\documentclass[twocolumn]{aastex631}

\newcommand\XVINtot{95}

\newcommand\XVImeanFeH{-2.17}
\newcommand\XVImeanFeHlerr{0.05}
\newcommand\XVImeanFeHuerr{0.05}

\newcommand\XVIdispFeH{0.33}
\newcommand\XVIdispFeHlerr{0.07}
\newcommand\XVIdispFeHuerr{0.07}

\newcommand\XVIskewFeH{-0.14}
\newcommand\XVIskewFeHlerr{0.24}
\newcommand\XVIskewFeHuerr{0.25}

\newcommand\XVIkurtFeH{0.21}
\newcommand\XVIkurtFeHlerr{0.38}
\newcommand\XVIkurtFeHuerr{0.53}

\newcommand\XVIgradRe{-0.23}
\newcommand\XVIdgradRe{0.15}

\newcommand\XVIgradkpc{-1.8}
\newcommand\XVIdgradkpc{0.9}

\newcommand\XVIgradamin{-0.27}
\newcommand\XVIdgradamin{0.14}

\newcommand\XVIinnerN{72}

\newcommand\XVImeaninnerFeH{-2.15}
\newcommand\XVImeaninnerFeHlerr{0.07}
\newcommand\XVImeaninnerFeHuerr{0.07}

\newcommand\XVIdispinnerFeH{0.37}
\newcommand\XVIdispinnerFeHlerr{0.08}
\newcommand\XVIdispinnerFeHuerr{0.08}

\newcommand\XVIskewinnerFeH{-0.17}
\newcommand\XVIskewinnerFeHlerr{0.26}
\newcommand\XVIskewinnerFeHuerr{0.27}

\newcommand\XVIkurtinnerFeH{0.15}
\newcommand\XVIkurtinnerFeHlerr{0.40}
\newcommand\XVIkurtinnerFeHuerr{0.55}

\newcommand\XVIouterN{23}

\newcommand\XVImeanouterFeH{-2.23}
\newcommand\XVImeanouterFeHlerr{0.10}
\newcommand\XVImeanouterFeHuerr{0.10}

\newcommand\XVIdispouterFeH{0.25}
\newcommand\XVIdispouterFeHlerr{0.14}
\newcommand\XVIdispouterFeHuerr{0.14}

\newcommand\XVIskewouterFeH{-0.09}
\newcommand\XVIskewouterFeHlerr{0.49}
\newcommand\XVIskewouterFeHuerr{0.48}

\newcommand\XVIkurtouterFeH{-0.13}
\newcommand\XVIkurtouterFeHlerr{0.6}
\newcommand\XVIkurtouterFeHuerr{0.96}

\newcommand\XXVIIINtot{191}

\newcommand\XXVIIImeanFeH{-1.95}
\newcommand\XXVIIImeanFeHlerr{0.04}
\newcommand\XXVIIImeanFeHuerr{0.04}

\newcommand\XXVIIIdispFeH{0.34}
\newcommand\XXVIIIdispFeHlerr{0.05}
\newcommand\XXVIIIdispFeHuerr{0.05}

\newcommand\XXVIIIskewFeH{-0.59}
\newcommand\XXVIIIskewFeHlerr{0.18}
\newcommand\XXVIIIskewFeHuerr{0.17}

\newcommand\XXVIIIkurtFeH{-0.76}
\newcommand\XXVIIIkurtFeHlerr{0.39}
\newcommand\XXVIIIkurtFeHuerr{0.49}

\newcommand\XXVIIIgradRe{-0.46}
\newcommand\XXVIIIdgradRe{0.10}

\newcommand\XXVIIIgradkpc{-1.39}
\newcommand\XXVIIIdgradkpc{0.44}

\newcommand\XXVIIIgradamin{-0.30}
\newcommand\XXVIIIdgradamin{0.10}

\newcommand\XXVIIIinnerN{144}

\newcommand\XXVIIImeaninnerFeH{-1.86}
\newcommand\XXVIIImeaninnerFeHlerr{0.04}
\newcommand\XXVIIImeaninnerFeHuerr{0.04}

\newcommand\XXVIIIdispinnerFeH{0.33}
\newcommand\XXVIIIdispinnerFeHlerr{0.07}
\newcommand\XXVIIIdispinnerFeHuerr{0.06}

\newcommand\XXVIIIskewinnerFeH{-0.52}
\newcommand\XXVIIIskewinnerFeHlerr{0.24}
\newcommand\XXVIIIskewinnerFeHuerr{0.23}

\newcommand\XXVIIIkurtinnerFeH{1.02}
\newcommand\XXVIIIkurtinnerFeHlerr{0.54}
\newcommand\XXVIIIkurtinnerFeHuerr{0.73}

\newcommand\XXVIIIouterN{47}

\newcommand\XXVIIImeanouterFeH{-2.28}
\newcommand\XXVIIImeanouterFeHlerr{0.11}
\newcommand\XXVIIImeanouterFeHuerr{0.10}

\newcommand\XXVIIIdispouterFeH{0.52}
\newcommand\XXVIIIdispouterFeHlerr{0.10}
\newcommand\XXVIIIdispouterFeHuerr{0.11}

\newcommand\XXVIIIskewouterFeH{-0.39}
\newcommand\XXVIIIskewouterFeHlerr{0.26}
\newcommand\XXVIIIskewouterFeHuerr{0.27}

\newcommand\XXVIIIkurtouterFeH{-0.15}
\newcommand\XXVIIIkurtouterFeHlerr{0.40}
\newcommand\XXVIIIkurtouterFeHuerr{0.55}

\shorttitle{Faint M31 Satellite Stellar Metallicities}
\shortauthors{Fu et al.} 

\begin{document}

\title{Stellar Metallicities and Gradients in the Faint M31 Satellites Andromeda~XVI and Andromeda~XXVIII}

\correspondingauthor{Sal Wanying Fu}
\email{swfu@berkeley.edu}

\author[0000-0003-2990-0830]{Sal Wanying Fu}
\affiliation{Department of Astronomy, University of California, Berkeley, Berkeley, CA, 94720, USA}

\author[0000-0002-6442-6030]{Daniel R. Weisz}
\affiliation{Department of Astronomy, University of California, Berkeley, Berkeley, CA, 94720, USA}

\author{Else Starkenburg}
\affiliation{Kapteyn Astronomical Institute, University of Groningen, Postbus 800, 9700 AV, Groningen, the Netherlands}

\author[0000-0002-1349-202X]{Nicolas Martin}
\affiliation{Universit\'{e} de Strasbourg, Observatoire astronomique de Strasbourg, UMR 7550, F-67000 Strasbourg, France}
\affiliation{Max-Planck-Institut f\"{u}r Astronomie, K\"{o}nigstuhl 17, D-69117 Heidelberg, Germany}

\author[0000-0002-1693-3265]{Michelle L. M. Collins}
\affiliation{School of Maths and Physics, University of Surrey, Guildford GU2 7XH, UK}

\author[0000-0002-1445-4877]{Alessandro Savino}
\affiliation{Department of Astronomy, University of California, Berkeley, Berkeley, CA, 94720, USA}

\author[0000-0002-9604-343X]{Michael Boylan-Kolchin}
\affiliation{Department of Astronomy, The University of Texas at Austin, 2515 Speedway, Stop C1400, Austin, TX 78712-1205, USA}

\author[0000-0003-1184-8114]{Patrick C\^{o}t\'{e}}
\affiliation{National Research Council of Canada, 
Herzberg Astronomy and Astrophysics Research Centre,
Victoria, BC V9E 2E7, Canada}

\author{Andrew E. Dolphin}
\affiliation{Raytheon, 1151 E. Hermans Road, Tucson, AZ 85756, USA}
\affiliation{Steward Observatory, University of Arizona, 933 North Cherry Avenue, Tucson, AZ 85721-0065 USA}

\author{Nicolas Longeard}
\affiliation{Laboratoire d’astrophysique, \'{E}cole Polytechnique F\'{e}d\'{e}rale de Lausanne (EPFL), Observatoire, 1290 Versoix, Switzerland}

\author{Mario L. Mateo}
\affiliation{Department of Astronomy, University of Michigan, 311 West Hall, 1085 S. University Avenue, Ann Arbor, MI 48109, USA}

\author[0000-0002-5908-737X]{Francisco J. Mercado}
\affiliation{Department of Physics and Astronomy, Pomona College, Claremont, CA 91711, USA}

\author[0000-0002-7393-3595]{Nathan R. Sandford}
\affiliation{Department of Astronomy and Astrophysics, University of Toronto, 50 St. George Street, Toronto ON, M5S 3H4, Canada}

\author[0000-0003-0605-8732]{Evan D. Skillman}
\affiliation{University of Minnesota, Minnesota Institute for Astrophysics, School of Physics and Astronomy, 116 Church Street, S.E., Minneapolis, MN 55455, USA}

\begin{abstract}
We present $\sim300$ stellar metallicity measurements in two faint M31 dwarf galaxies, Andromeda~XVI ($M_V = -7.5$) and Andromeda~XXVIII ($M_V = -8.8$) derived using metallicity-sensitive Calcium H~\& K narrow-band Hubble Space Telescope imaging. These are the first individual stellar metallicities in And~XVI (95 stars). Our And~XXVIII sample (191 stars) is a factor of $\sim15$ increase over literature metallicities. For And~XVI, we measure $\langle \mbox{[Fe/H]}\rangle = \XVImeanFeH^{+\XVImeanFeHuerr}_{-\XVImeanFeHlerr}$, $\sigma_{\mbox{[Fe/H]}}=\XVIdispFeH^{+\XVIdispFeHuerr}_{-\XVIdispFeHlerr}$, and $\nabla_{\mbox{[Fe/H]}}=\XVIgradRe \pm \XVIdgradRe$~dex~$R_e^{-1}$. We find that And~XVI is more metal-rich than MW UFDs of similar luminosity, which may be a result of its unusually extended star formation history. For And~XXVIII, we measure $\langle \mbox{[Fe/H]}\rangle = \XXVIIImeanFeH^{+\XXVIIImeanFeHuerr}_{-\XXVIIImeanFeHlerr}$, $\sigma_{\mbox{[Fe/H]}}=\XXVIIIdispFeH^{+\XXVIIIdispFeHuerr}_{-\XXVIIIdispFeHlerr}$, and $\nabla_{\mbox{[Fe/H]}}=\XXVIIIgradRe \pm \XXVIIIdgradRe$~dex~$R_e^{-1}$, placing it on the dwarf galaxy mass-metallicity relation. Neither galaxy has a metallicity distribution function with an abrupt metal-rich truncation, suggesting that star formation fell off gradually. The stellar metallicity gradient measurements are among the first for faint ($L \lesssim 10^6~L_{\odot}$) galaxies outside the Milky Way halo. Both galaxies' gradients are consistent with predictions from the FIRE simulations, where an age-gradient strength relationship is the observational consequence of stellar feedback that produces dark matter cores. We include a catalog for community spectroscopic follow-up, including 19 extremely metal poor ($\mbox{[Fe/H]}<-3.0$) star candidates, which make up 7\% of And~XVI's MDF and 6\% of And~XXVIII's. 
\end{abstract}
\keywords{Dwarf galaxies (416); HST photometry (756); Local Group (929); Stellar abundances (1577)}

\section{Introduction} \label{sec:intro}

\par Studies of satellite galaxies around the next-nearest spiral galaxy, M31, have proceeded alongside studies of Milky Way (MW) satellite galaxies over the last two decades. The story of M31 satellite discovery is familiar: just as advances in photometric surveys have unearthed many more MW satellites (e.g., \citealt{belokurov2007sdss}, \citealt{bechtol2015DESsat}, \citealt{drlicawagner2015}, \citealt{laevens2015discovery}, \citealt{homma2019}), deep and comprehensive photometric surveys have similarly built an abundant census of faint satellites around M31 (e.g., \citealt{mcconnachie2006m31dwarfsstructural}, \citealt{mcconnachie2008m31sats}, \citealt{martin2013m31dwarfsearch}, \citealt{dolivadolinsky2022pandasiii}). 

\par Emerging studies of these satellites indicate that their properties are not entirely similar to those of MW dwarf galaxies (\citealt{mcconnachie2006m31satellitedistr}, \citealt{brasseur2011glxsize}, \citealt{collins2014dwarfmassprofile}, \citealt{savino2023}, \citealt{dolivadolinsky2023pandasiv}). Most notably in the ultra-faint regime ($<10^5~L_{\odot}$), their SFHs tend to be more extended than MW ultra-faint dwarf galaxies (UFDs), suggesting that the low-mass galaxies are not uniformly affected by reionization (e.g., \citealt{weisz2014m31sfhs}, \citealt{monelli2016and16}, \citealt{savino2023}). The distinct SFHs of M31 satellites suggest that they experienced different chemical enrichment histories than MW UFDs. The evidence of this difference should be reflected in their stellar abundance patterns (e.g., distributions in [Fe/H], [$\alpha$/Fe]), which encode information the baryonic processes (e.g., the balance of inflows vs outflows) that shaped a galaxy's evolution (e.g., \citealt{andrews2017chemevmodels}, \citealt{weinberg2017chemev}, \citealt{sandford2022eriII}).  Unfortunately, the large distance of the M31 satellites means that only the brightest have ample robust stellar abundances measured from spectroscopy (\citealt{ho2012and2}, \citealt{vargas2014m31alpha}, \citealt{kvasova2024and18}). In fainter M31 satellites, only a handful of individual stars are bright enough for ground-based measurements \citep{kirby2020m31dwarfs}.

\par Photometric metallicity measurement techniques enable recovery of stellar metallicities for fainter stars lower down a galaxy's luminosity function by imaging in filters specifically designed to trace strong, metallicity-sensitive features in stellar spectra. They are an important complement to ongoing spectroscopic efforts to study the M31 dwarf population in great detail. This is a well-established technique that has been utilized for decades of MW studies, including to the present (e.g., \citealt{stromgren1966}, \citealt{beers1985cahk}, \citealt{starkenburg2017pristine}, \citealt{chiti2021skymapper}, \citealt{martin2023pristine}). From space, our team has already applied this imaging technique using HST/WFC3 imaging in the narrow-band $F395N$ filter, targeting the Ca H\&K lines, to great success for Milky Way UFDs (\citealt{fu2022eriII,fu2023ufds}), and to a distant isolated dwarf galaxy on the outskirts of the Local Group (Tucana; \citealt{fu2024tucana}). The singular and singularly important contribution HST has made to the field of resolved stellar metallicity measurements is enabling measurements for stars 1-2 magnitudes fainter than currently reachable by any ground-based spectroscopic or imaging facility to-date. 

\par In this paper, we present results from an HST program to apply the narrow-band CaHK photometric imaging technique to two of M31's UFD satellite galaxies: Andromeda XVI (And~XVI) and Andromeda XXVIII (And~XXVIII). 

\par And~{\sc XVI} is a rare system among the $\sim100$ cataloged low-mass galaxies in the Local Group \citep{ibata2007m31photosurvey}. Its stellar mass ($M_V=-7.5$, $M_{\star}\sim10^5 M_{\odot}$) places it among the brightest ``ultra-faint'' dwarf galaxies (UFDs) known in the LG \citep{simon2019review}. At $280$~kpc from M31 \citep{savino2022}, And~{\sc XVI} lies beyond the spiral's virial radius. While the vast majority of UFDs are characterized by metal-poor, ancient stellar populations that are associated with formation at times during or prior to reionization, deep HST imaging revealed that And~{\sc XVI} only had a low-level of star formation at the earliest times, and instead experienced, extended star formation that lasted until $\sim6$~Gyr ago ($z\lesssim1$; \citealt{weisz2014m31sfhs}, \citealt{monelli2016and16}, \citealt{skillman2017islands}). And~{\sc XVI}'s prolonged SFH provided a new link between the faintest UFDs that appear to be quenched by reionization and more massive ``classical'' dwarfs whose growth appears unaffected by reionization. Detailed stellar MDFs are a crucial component of characterizing the physics driving these remarkable properties, but prior to this work, there have been no resolved stellar metallicity measurements in And~XVI: Keck/DEIMOS spectroscopy was only able to obtain mean metallicity measurements $\langle \mbox{[Fe/H]} \rangle = -2.1 \pm 0.2, -1.9 \pm 0.2,$ and $-2.0 \pm 0.1$ by stacking low-S/N stellar spectra (\citealt{letarte2009deimosm31sats}, \citealt{collins2013m31dwarfs}, \citealt{collins2015m31plane}).

\par And~XXVIII ($M_{\star} \sim 10^{5.5}~M_{\odot}$) is a faint galaxy located $\sim370$~kpc from the center of M31 \citep{slater2011and28,savino2022}. More massive than And~XVI, its SFH is also more strongly weighted to old ages: the majority of its stars formed 12~Gyr ago, with a small remaining fraction forming as recently as 8~Gyr ago; these SFH properties are comparable to that of the Ursa Minor and Draco dSphs \citep{skillman2017islands}. Using Keck/DEIMOS spectroscopy, \citet{slater2015spectroscopy} measured Calcium triplet metallicities for thirteen stars in And~XXVIII, finding $\langle \mbox{[Fe/H]} \rangle = -1.84 \pm 0.15$~dex and $\sigma_{\mbox{[Fe/H]}} = 0.65 \pm 0.15$~dex. While these measurements are illustrative for tracing out broad properties of its MDF, they are not well-sampled enough to fully trace out its shape and define the tails.

\par Studying just these two galaxies can offer insight on star-forming physics in dwarf galaxies of different mass and over different timescales in an environment that is not the MW halo. Robust stellar metallicities are a critical missing piece to this emerging picture. The properties of And~XXVIII and And~XVI formed the basis for an approved pilot HST program leveraging Ca H\&K imaging on WFC3/UVIS to measure their stellar MDFs. In this paper, we present the first extensive set of resolved stellar metallicities in And~XVI and And~XXVIII. We present our observations in \S \ref{sec:obsdata}, our metallicity inference method in \S \ref{sec:metmethods}, and our results in \S \ref{sec:results}. We discuss the implications of our results in \S \ref{sec:discussion}, and conclude with forward-facing remarks in \S \ref{sec:conclusion}.

\section{Observations \& Data Reduction} \label{sec:obsdata}

\begin{figure*}
    \epsscale{1.2}
    \centering
\includegraphics[scale=0.35]{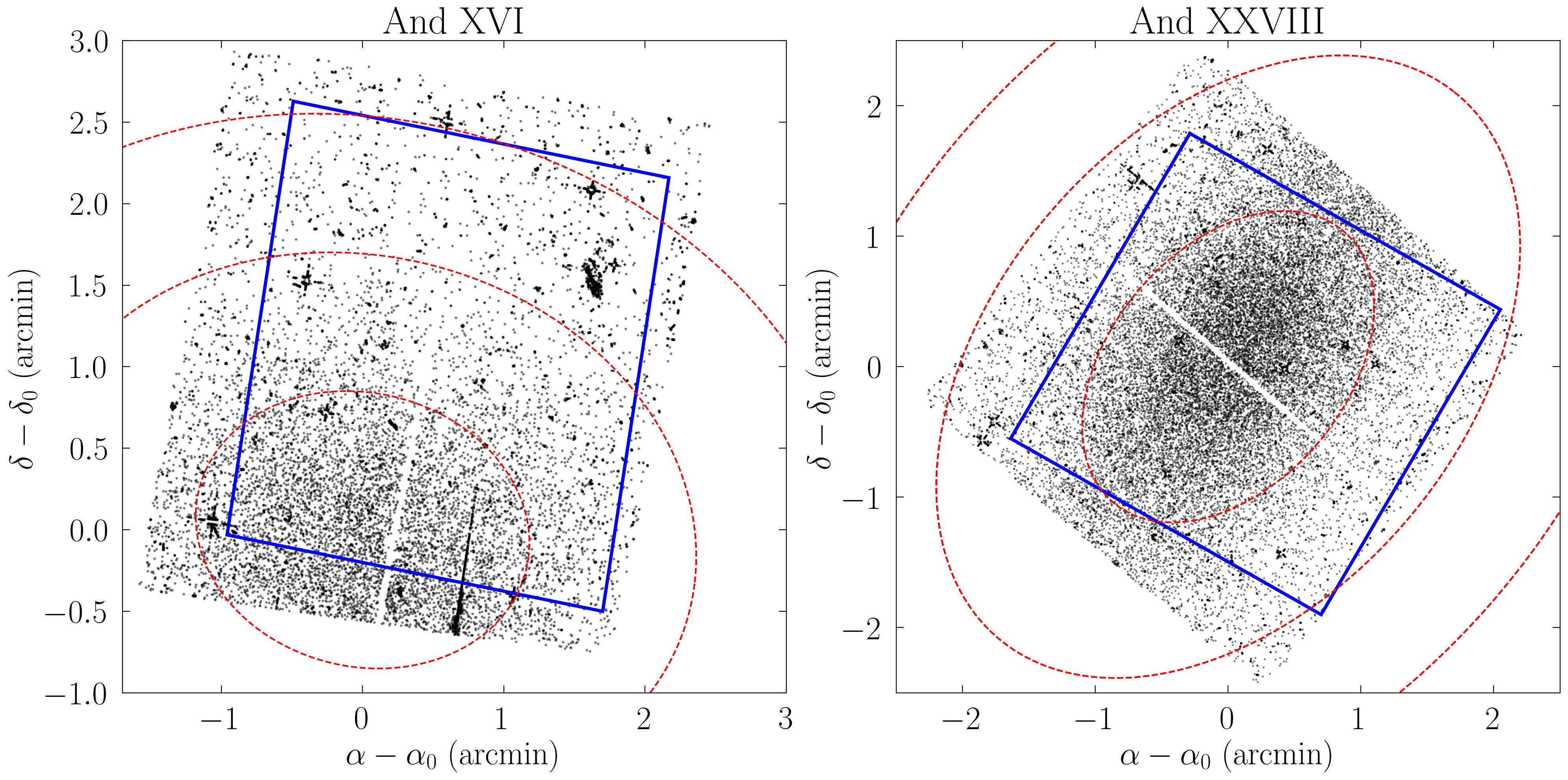}
    \caption{Illustrating the coverage of our HST imaging. The black dots are detected sources from the archival ACS archival image for each galaxy from which we measure broadband $F475W$ and $F814W$ photometry. The blue squares are the WFC3/$F395N$ FoV. Each galaxy's elliptical half-light contours are plotted in red, corresponding to 1, 2, and 3 half-light elliptical radii. }
    \label{fig:cameraFoVs}
\end{figure*}

\begin{deluxetable*}{cCcCc}
\tablecaption{Dwarf Galaxy Characteristics}
\tablehead{\colhead{Parameter} & \colhead{And~XVI} & \colhead{And~XVI ref.} & \colhead{And~XXVIII} & \colhead{And~XXVIII ref.}}
\startdata
R.A. (deg.)                          & 14.8762500     & \citet{martin2016m31sats} &   338.1716667 & \citealt{slater2011and28} \\
Dec. (deg.)                          & 32.3761111     & ''                       &    31.2161667 & ''                   \\
Ellipticity                          & 0.29 \pm 0.08  & ''                       & 0.34 \pm 0.13 & ''                       \\
P.A. (deg.)                          & 99 \pm 9       & ''                       &     39 \pm 16 & ''                       \\
$r_h$ ($\arcmin$)                    & 1.0 \pm 0.1    & ''                       & 1.11 \pm 0.21 & ''                       \\
$E(B-V)$                             & 0.0568         & \citet{schlafly2011}      & 0.0758        & \citet{schlafly2011} \\ 
\hline
$(m - M)_0$ (mag.)                   & 23.57 \pm 0.08 & \citet{savino2022}        & 24.36 \pm 0.05  & \citet{savino2022} \\
$M_V$ (mag.)                         & -7.5 \pm 0.03  & ''                       & -8.8 $^{+0.4}_{-1.0}$         & ''   \\    
Luminosity (log$_{10} L_{\odot}$)    & 4.9            & ''                       & 5.5           & ''                       \\ 
$M_{\star}$ (log$_{10} M_{\odot}$)   & 5.2            & ''                       & 5.8           & ''                    \\
$D_{\odot}$ (kpc)                    & 517 \pm 19   &  ''                      & 745 \pm 17        & ''                \\
$D_{M31}$ (kpc)                      & 280^{+26}_{-27} &  ''                 & 368.8^{+7.8}_{-7.3} & ''                \\
\hline
$r_h$ in WFC3 FoV                    & 2.7               & This work                & 2.43                & This work           \\
F475W exp. time (s)                  & 17194             & \citet{skillman2017islands} & 26360               & \citet{skillman2017islands} \\ 
F814W exp. time (s)                  & 13622             & ''                      & 20880               & '' \\ 
F395N exp. time (s)                  & 30154             & This work                & 24869               & This work           \\ 
F395N obs. dates                     & \mbox{Jan. 9-15, 2023}   & ''                      & \mbox{Nov. 21-23, 2022}    & ''     \\ 
\enddata
\tablecomments{Observational characteristics of And~XVI and And~XXVIII. $F395N$ data are taken for this work from HST GO-16686 (PI: Weisz). Archival broadband data are from the ISLAndS program \citep{skillman2017islands}: GO-13028 (PI: Skillman) for And~XVI \citep{weisz2014sfh,monelli2016and16,skillman2017islands}, and GO-13739 (PI: Skillman) for And~XXVIII.}
\label{tab:dwarfparams}
\end{deluxetable*}

\par We acquired new F395N imaging using HST WFC3/UVIS for And~{\sc XVI} and And~{\sc XXVIII} as part of HST-GO-16686 (PI: Weisz).  Our aim was to measure metallicities for $\sim100$ red giant stars in each system.  Accordingly, the nominal  target depth of this program aimed to collect sufficiently high SNR ($\gtrsim10$) F395N imaging to the approximately the depth of the HB ($M_{\rm F475W} \sim +0.5$). This depth goal is slightly brighter in the more luminous systems, as reaching the HB was not necessary to acquire at least 100 stars.

Imaging was acquired between November 21, 2022 and January 15, 2023. We summarize our observations in Table \ref{tab:dwarfparams}. We also targeted And~{\sc XV}, but the observations, and the repeat observations, failed due to issues with guide star acquisition. 

\par We centered our UVIS fields on existing HST/ACS broadband F475W and F814W imaging from the ISLAnds program \citep{skillman2017islands}. We used sub-pixel dithers to sample the point spread function, improve cosmic ray rejection, etc.\ and used the post-flash level (\texttt{FLASH}$=20$) advised by STScI to mitigate CTE effects. We observed And~XI for 11 orbits and And~XXVIII for 9 orbits.  Visits for both sets of observations consisted of 1-2 orbits.  Long integrations were used in order to increase the S/N for bright stars in order to aid with astrometric alignment of the images.  More observational details can be found in our public Phase II file. Figure \ref{fig:cameraFoVs} shows the WFC3 footprint overlaid on the ACS image of each galaxy, overplotted with elliptical contours at 1, 2, and 3 half-light radii. 

\par We performed point spread function (PSF) photometry simultaneously on all F395N, F475W, and F814W HST images of And~{\sc XVI} and And~{\sc XXVIII} using DOLPHOT (\citealt{dolphin2000hstphot}, \citealt{dolphot2016}).  DOLPHOT is a crowded field stellar photometry package that is widely used to analyze resolved star observations of nearby galaxies and star clusters.  We followed the DOLPHOT reduction procedure that our team has applied to HST Ca H\&K studies of other nearby galaxies (\citealt{fu2022eriII, fu2023ufds, fu2024tucana}).

\par Following \citet{fu2023ufds}, we culled the raw DOLPHOT output in order to select high quality stars.  We created a noise model for our data using $\sim500,000$ artificial star tests (ASTs).  The ASTs were distributed across the red giant branch (RGB) in F395N, F475W, and F814W, following the method described in \citet{fu2024tucana}.  

\par
\begin{figure*}
    \epsscale{1.2}
    \centering
\includegraphics[scale=0.3]{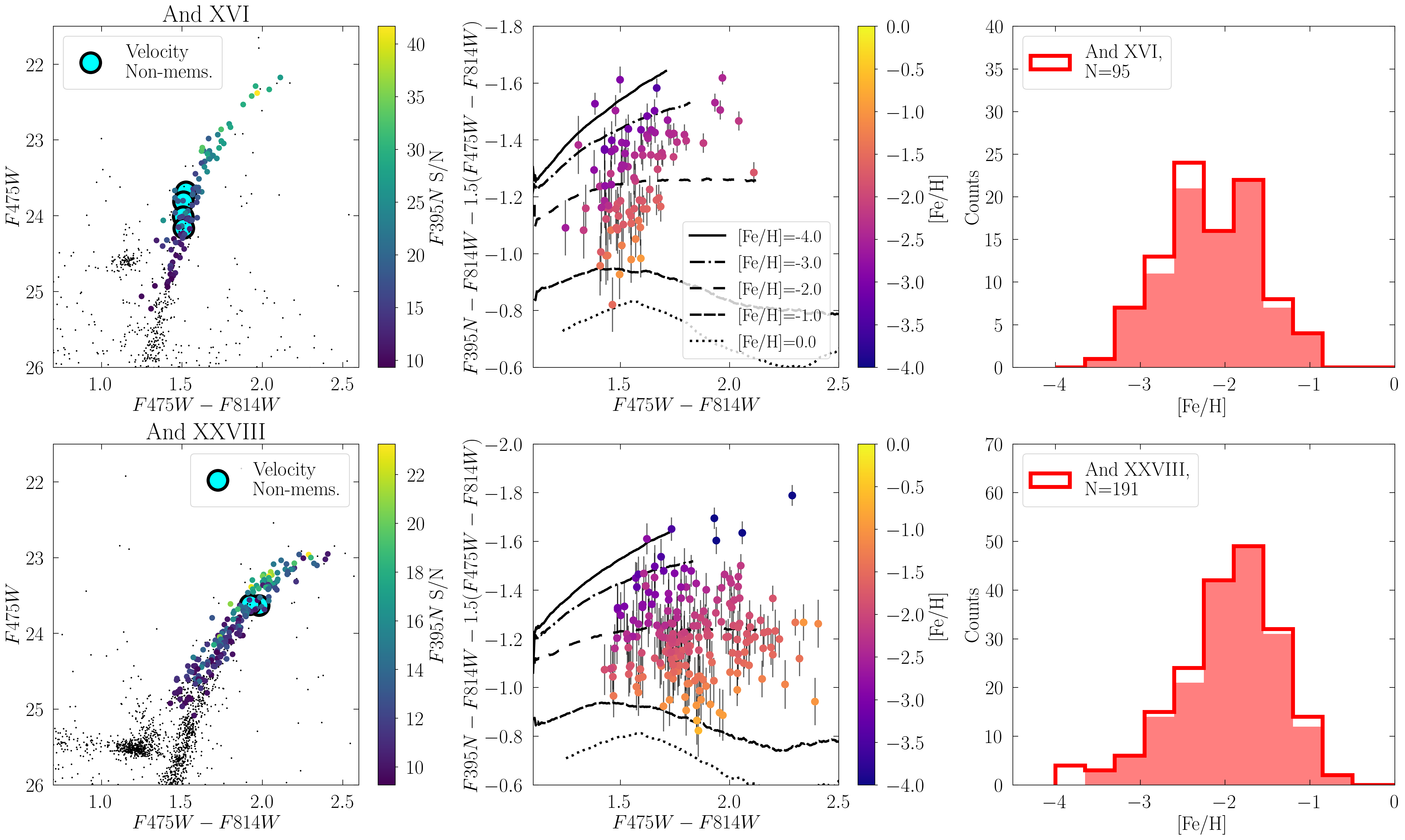}
    \caption{ Presenting the CMD, CaHK color data, and resulting MDF for And~XVI \textbf{(top row)} and And~XXVIII \textbf{(bottom row)}. \textbf{Left column:} Broadband CMD of each galaxy, centered on its RGB, with stars analyzed color-coded by their $F935N$ S/N. Stars in gray triangles are velocity interlopers determined using the datasets of \citet{collins2013m31dwarfs,collins2014dwarfmassprofile,collins2015m31plane}. \textbf{Center column:} The position of analyzed stars in the \textit{Pristine}-like color-color space, color-coded by their inferred metallicities. Mono-metallic CaHK tracks, convolved by the AST profile of a star of median luminosity in each galaxy's sample, are overplotted to guide the eye. Some of these metallicities are inferred upper limits due to either falling at the edge of the model grid for bright stars, or being low S/N for fainter stars. \textbf{Right column:} The resulting MDF for each galaxy. Solid histograms are well-constrained measurements, and empty histograms are stars for which we can only derive an upper limit. }
    \label{fig:cahkmdf}
\end{figure*}

 The left column of Figure \ref{fig:cahkmdf} shows the broadband $F475W-F814W$ CMDs of And~{\sc XVI} (top) and And~{\sc XXVIII} (bottom), zoomed-in on their RGBs. We use the kinematic catalogs of each dwarf used in \citet{collins2013m31dwarfs, collins2014dwarfmassprofile, collins2015m31plane}, to remove 2 radial velocity interlopers in And~XXVIII and 4 interlopers in And~XVI for stars as faint as $F475W \sim 24$; these stars are represented in the figure as shaded triangles. We color-code the RGB stars used in this analysis by their F395N signal-to-noise. The F395N data reached the anticipated depth (i.e., approximately the HB) in both cases. Following \citet{fu2023ufds}, we only measure metallicities for stars with $SNR_{\rm F395N} \ge 10$. After measuring metallicities, we remove additional obvious interlopers by hand; namely, stars whose metallicities are inconsistent with their color on the CMD (e.g., metal-rich stars on the blue end of the RGB). This resulted in the removal of an additional 3 stars in And~XXVIII and 4 stars in And~XVI. These vetting results are within the range of predicted MW foreground contamination from the TRILEGAL model \citep{vanhollebeke2009trilegal}, which we expect to have a minimal contribution to the MDFs due to the small FoV of HST \citep{fu2023ufds}. 

\par In total, our sample includes \XXVIIINtot~stars in And~XXVIII and \XVINtot~stars in And~XVI. 

\section{Metallicity Property Determination} \label{sec:metmethods}

\subsection{Individual Stellar Metallicities} \label{sec:indivmethods}

\par We measure stellar metallicities following the procedures laid out in \citet{fu2023ufds}, with adaptations adopted in \citet{fu2024tucana} for the case of dwarf galaxies with extended SFHs. Here, we briefly summarize this technique.

\par We construct a set of basis functions from the MESA Stellar Isochrones and Tracks (MIST) stellar evolution models \citep{choi2016mist,dotter2016mist} over a range of metallicities ($-4\le$[Fe/H]$\le0$) for a 13~Gyr stellar population. These models are the forthcoming v2 models which are scaled to the \citet{grevesse1998} solar abundances, and include variations in [$\alpha$/Fe]. Because the RGB is only weakly sensitive to age, the exact age adopted does not affect Ca H\&K metallicity measurements.  The basis functions are projected into the standard Ca H\&K color-color space, as shown in the middle panels of Figure \ref{fig:cahkmdf}. The color-space is pixelated (i.e., similar to a Hess diagram used in fitting CMDs). We adopt bins that are 0.025 in size in both color indices.  Each  basis function is then convolved with the noise model determined from the ASTs.  We compare the location of each observed star in Ca H\&K color-color space to each basis function, using a Poisson likelihood function and flat priors in [Fe/H].   

\par We follow \citet{fu2024tucana} to account for the effects of varying $\alpha$-elements. Specifically, for [Fe/H]$<-2.0$, we only use basis functions with [$\alpha$/Fe]$=+0.4$ and for [Fe/H]$>-1.0$ we only use basis functions with [$\alpha$/Fe]$=+0.0$. Finally, for the range $-2.0\le$[Fe/H]$\le-1.0$, we use basis functions with [$\alpha$/Fe]$=+0.2$. This criteria approximates the plateau, knee, and knee-to-ankle transitions known to exist in the [$\alpha$/Fe] vs [Fe/H] diagrams of Milky Way satellites (e.g., \citealt{tolstoy2009dsphreview}). As demonstrated in \citet{fu2022eriII}, the choice of [$\alpha$/Fe] introduces a shift in metallicity measurements.  For example, adopting [$\alpha$/Fe]$=+0.4$ instead of [$\alpha$/Fe]$=+0.0$ causes the measured metallicity to be $\sim0.2$~dex more metal-rich.  In the current analysis technique, we do not consider any scatter in [$\alpha$/Fe] at fixed [Fe/H].  In addition to random uncertainties on each star, we adopt a 0.2~dex systematic uncertainty owing to uncertainties in  [$\alpha$/Fe] and the impact of additional elements (e.g., \citealt{fu2023ufds}).

\par We evaluate the posterior distribution for each star using \texttt{emcee} \citep{emcee}. We use 50 walkers with a burn-in time of 50 steps per star, and then sample of $10^4$ steps to fill out the posterior space. We assess convergence using the Gelman-Rubin statistic (\citealt{gelmanrubin1992}).

\par Following past work, for stars with PDFs that have clear peaks and are within the metallicity grid, we report the median of each PDF as the metallicity of the star, and the 68\% confidence interval for the statistical uncertainty. The statistical uncertainty is set by spacing in the MIST CaHK color tracks, and photometric uncertainty. We sum the statistical uncertainties in quadrature with an additional 0.2~dex systematic uncertainty in RGB star metallicities, as determined by \citet{fu2023ufds}. The handful of extremely metal-poor stars in our sample (e.g., [Fe/H] $\le -3.0$), that have either well- or poorly-constrained PDFs (i.e. severe truncation at the grid boundary), tend to have larger statistical uncertainties owing to the closer spacing (i.e., decreased sensitivity) of Ca H\&K tracks at lower metallicity. However, a star with very high photometric S/N can have a well constrained metallicity below [Fe/H] $\le -3.0$. For these stars, we increase the value of the systematic uncertainty to 0.5~dex to reflect challenges in accurately modeling metal-poor stars. 

\subsection{MDF Summary Statistics}

\par We compute summary statistics (e.g., fit a Gaussian profile) for the MDFs of And~XVI and And~XXVIII, and spatial subpopulations, following the methods described in \citet{fu2024tucana} and references therein. We compute the mean metallicity ($\langle$[Fe/H]$\rangle$) and metallicity dispersion ($\sigma_{\mbox{[Fe/H]}}$) of the MDF by using a two-parameter Gaussian likelihood function. We assume a uniform prior on mean metallicity bounded by the most MP and MR star in the MDF. We also require the metallicity dispersion to be greater than zero. Similar to previous procedure, we adopt symmetric Gaussian uncertainties on individual measurements. We sample the posterior distribution using \texttt{emcee} \citep{emcee}, initializing 50 walkers to run for 10,000 steps. The autocorrelation time is 50 steps, and we assess convergence using the GR statistic. 

\par For higher-order statistics such as skew and kurtosis, we also assume symmetric Gaussian uncertainties on individual metallicitiy measurements, and use them to construct 10,000 realizations of the MDF. We measure the skew and kurtosis from each realization. We report final skew and kurtosis measurements as the median of the overall distribution, with lower and upper uncertainties set by the 16th and 84th percentiles. We present our table of summary statistics in Table \ref{tab:metprops}, and our table of individual stellar metallicities in Table \ref{tab:indivmeasurements}. 

\subsection{Measuring Metallicity Gradients}

\par The spatial extent of our imaging enables us to characterize spatial metallicity trends across each galaxy. Thus we also quantify the strength of the gradient as the slope ($\nabla_{\mbox{[Fe/H]}}$) obtained by fitting a line to individual stellar metallicities as a function of elliptical $R_e$ from the center. Specifically, we follow the procedure outlined in \citet{hogg2010datarecipes} and implemented in \citet{fu2024tucana}, which assumes Gaussian uncertainties on our measurements. We also fit the intercept of the line ($\mbox{[Fe/H]}_0$) and marginalize over an additional parameter $f$ that is the fractional underestimation of measurement uncertainties. We assume uniform priors over the slope, intercept, and logarithmic fractional uncertainty. We sample the distribution using \texttt{emcee} by running 32 walkers for 10,000 steps. The burn-in time is about 50 steps. We assess convergence using the GR statistic. 

\section{Results} \label{sec:results}

\par We present the results of our MDF measurements. The center column of Figure \ref{fig:cahkmdf} presents our sample in the color-color space defined by broadband color $F475W-F814W$ and the \textit{Pristine}-like color index (\citealt{starkenburg2017pristine}; $F395N - F475W - 1.5(F475W - F814W)$). Table \ref{tab:metprops} summarizes our global metallicity measurements. We present individual metallicities in Table \ref{tab:indivmeasurements}, and candidate EMP stars in Table \ref{tab:EMPstars}. In the subsequent sub-sections, we discuss results for individual galaxies.  

\begin{figure*}
    \epsscale{1.2}
    \centering
\includegraphics[scale=0.45]{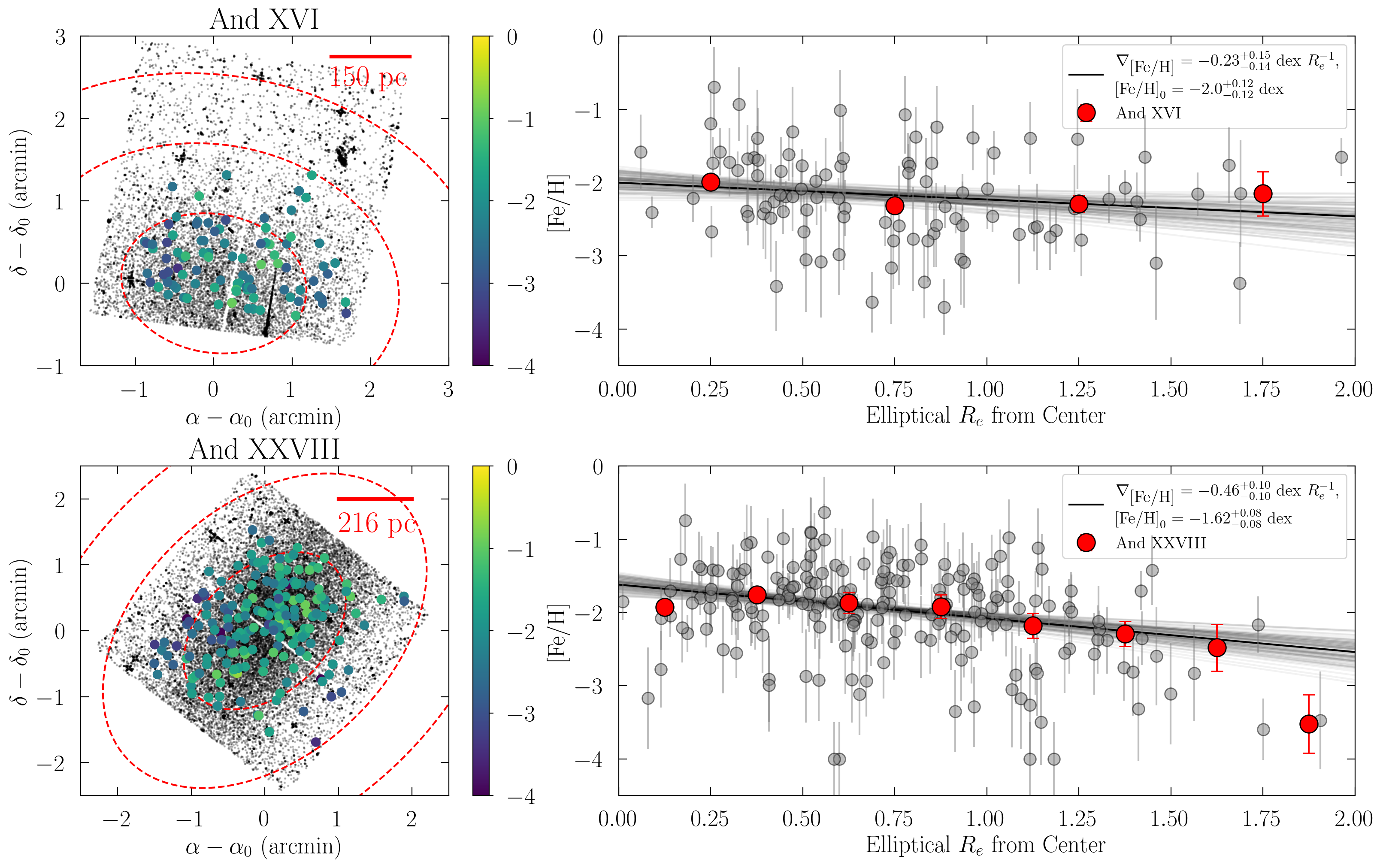}
    \caption{\textbf{(Left)} Spatial distribution of stars in And~XVI \textbf{(top)} and And~XXVIII \textbf{(bottom)}, color-coded by their inferred metallicities. Also presented are each galaxy's half-light contours, and a scale bar indicating physical distance. \textbf{(Right)} Result of fitting a linear model to our data as a function of elliptical half-light radius. Gray dots are individual data points in each galaxy, while red dots are the mean metallicity in spatial bins of 0.5~$R_e$ for And~XVI and 0.25~$R_e$ for And~XXVIII. We are able to robustly recover a metallicity gradient for And~XXVIII. We also tentatively detect a gradient in And~XVI. These are among the few metallicity gradients of M31 satellites more generally, and for And~XVI, one of few gradients detected in UFDs to-date. }
    \label{fig:gradientsspatial}
\end{figure*}

\subsection{And~XVI} \label{sec:resultsand16}

\par The top row of Figure \ref{fig:cahkmdf} shows the broadband CMD of our member stars (left), their position on the CaHK color space with a representative set of AST-convolved CaHK mono-metallic tracks (center), and the resulting MDF constructed from measurements of \XVINtot\ stars (right). The stars in our MDF span a range from $-3.5$ to $-1$, with the majority of them between $-2.5$ and $-1.5$ (57 stars, 60\%). Eleven (12\%) stars have metallicities above $-1.5$. Seven stars (7\%) have metallicities below $-3$; we designate these stars as EMP candidates. We measure an overall mean metallicity of $\langle \mbox{[Fe/H]} \rangle = \XVImeanFeH^{+\XVImeanFeHuerr}_{-\XVImeanFeHlerr}$, and quantify higher-order moments in the MDF. We measure a metallicity dispersion of $\sigma_{\mbox{[Fe/H]}} = \XVIdispFeH^{+\XVIdispFeHuerr}_{-\XVIdispFeHlerr}$. We also quantify the skew and kurtosis, but do not find deviations from Gaussianity at a $2\, \sigma$ significance. Visually, there appears to be a double peak in its MDF (e.g. Figure \ref{fig:cahkmdf}) at $\sim -2.5$ and $\sim -1.7$, though the peaks may become washed out by moving a handful of stars within uncertainty into an adjacent bin. 

\par To-date, mean metallicity determinations of And~XVI are from co-addition of S/N $\gtrsim 3$ spectra in the Ca II triple line regime (\citealt{collins2015m31plane}, \citealt{letarte2009deimosm31sats}, using the \citealt{starkenburg2010catlomwet} calibration), which finds a mean metallicity of $\langle \mbox{[Fe/H]} \rangle = -2.0 \pm 0.1$. Our mean metallicity is in good agreement with this coarser determination, and places And~XVI within 1~$\sigma$ scatter of the dwarf galaxy mass-metallicity relation \citep{kirby2013LZR}.

\par We measure the metallicity gradient in And~XVI to $2~R_e$. In the top row of Figure \ref{fig:gradientsspatial}, 
we show the on-sky spatial distribution of stars, color-coded by our measured metallicity, and their distribution as a function of $R_e$. 
We measure a gradient of $\nabla_{\mbox{[Fe/H]}} = \XVIgradRe \pm \XVIdgradRe$~dex~$R_e^{-1}$. We consider the implications of this measurement as tentative evidence for a metallicity gradient and discuss its implications further in Section \ref{sec:discussion}.

\subsection{And~XXVIII} \label{sec:resultsand28}

\par We present our sample and the resulting inferred MDF in the bottom row of Figure \ref{fig:cahkmdf}. The stellar metallicities span a range from as low as $-4$ to as metal-rich as $-0.5$. We identify 12 (6\%) stars below $-3$ that we flag as EMP candidates for spectroscopic followup. Eight stars (4\%) of stars are more metal-rich than $-1.0$. The MDF is peaked at $\sim-1.8$, and majority of the stars (101; 53\%) are between $-2.0$ and $-1.0$. We measured a mean metallicity of $\langle \mbox{[Fe/H]} \rangle = \XXVIIImeanFeH^{+\XXVIIImeanFeHuerr}_{-\XXVIIImeanFeHlerr}$ and dispersion $\sigma_{\mbox{[Fe/H]}}= \XXVIIIdispFeH^{+\XXVIIIdispFeHuerr}_{-\XXVIIIdispFeHlerr}$. We also find that the MDF of And~XXVIII is skew-negative above $3\sigma$ significance- this is reflected in the MDF, which shows a long MP tail. 

\par Our results expand significantly on \citet{slater2015spectroscopy}, who measured individual metallicities for 13 And~XXVIII stars using Gemini/GMOS spectroscopy. They measure stellar metallicities using the \citet{carrera2013cat} Calcium Triplet (CaT) calibration and from Gaussian equivalent widths of the strongest two CaT lines. Their stars span a range similar to ours: $-3.0$ and $-0.5$. Due to their sparse number sampling, the stars are evenly distributed across this range. From this dataset, they measured $\langle \mbox{[Fe/H]} \rangle = -1.84 \pm 0.15$, which is in good agreement with our mean metallicity measurement. Their metallicity dispersion measurement of $0.65 \pm 0.15$ disagrees with our measurement by $2\sigma$. This discrepancy is in part due to the improved sampling of our MDF about the mean. Spatial sampling also matters here, given the strong metallicity gradient in And~XXVIII: the broader spatial coverage of the \citet{slater2015spectroscopy} dataset gives MP stars more weight in the MDF, therefore broadening the dispersion.

\par Of the 7 stars that we have in common with \citet{slater2015spectroscopy}, our metallicities appear to be more MR by $\sim~1$~dex, but this disagreement is also within their uncertainties ($\sim0.5-1.0$~dex). The formal level of agreement between our measurements is $1\ \sigma$. As shown in Figure 6 of \citet{fu2024tucana}, our CaHK metallicities are generally in agreement with CaT metallicities from the literature which themselves are derived from different implementations of the CaT calibration. 

\par We resolve a strong gradient for And~XXVIII of $\nabla_{\mbox{[Fe/H]}}=\XXVIIIgradRe \pm \XXVIIIdgradRe$~dex~$R_e^{-1}$. We present the spatial distribution of stars and their metallicity as a function of $R_e$ in Figure \ref{fig:gradientsspatial}. Here, the spatial concentration of metal-rich stars in comparison to metal-poor stars is apparent even by eye. We attempted to formally measure the scale radii of the metal-poor and metal-rich stars, but did not find a difference-- this is likely because our data do not span sufficient area across the galaxy. Our measurement shows that there is a $\sim0.9$~dex difference in mean metallicity between the population at the center of the galaxy and the population out at $2~R_e$, across $\sim430$~pc. We explore the implications of this gradient more in \S \ref{sec:discussion}. 

\begin{deluxetable*}{ccCC}
\label{tab:metprops}
\tablecaption{Metallicity Properties of And~XVI and And~XXVIII}
\tablehead{\colhead{Feature} & \colhead{Parameter} & \colhead{And~XVI} & \colhead{And~XXVIII}}
\startdata
Global MDF & $N$                        & \XVINtot                                          & \XXVIIINtot \\
           & $\langle$[Fe/H]$\rangle$ & \XVImeanFeH^{+\XVImeanFeHuerr}_{-\XVImeanFeHlerr} & \XXVIIImeanFeH^{+\XXVIIImeanFeHuerr}_{-\XXVIIImeanFeHlerr} \\
           & $\sigma_{\mbox{[Fe/H]}}$   & \XVIdispFeH^{+\XVIdispFeHuerr}_{-\XVIdispFeHlerr} & \XXVIIIdispFeH^{+\XXVIIIdispFeHuerr}_{-\XXVIIIdispFeHlerr} \\
           & Skew                       & \XVIskewFeH^{+\XVIskewFeHuerr}_{-\XVIskewFeHlerr} & \XXVIIIskewFeH^{+\XXVIIIskewFeHuerr}_{-\XXVIIIskewFeHlerr} \\
           & Kurtosis                   & \XVIkurtFeH^{+\XVIkurtFeHuerr}_{-\XVIkurtFeHlerr} & \XXVIIIkurtFeH^{+\XXVIIIkurtFeHuerr}_{-\XXVIIIkurtFeHlerr} \\ 
\hline
Metallicity Gradient   & $R_e$ Extent                              & 2                          & 2                                 \\
                       & $\nabla_{\mbox{[Fe/H]}}$ (dex~$R_e^{-1}$) & \XVIgradRe \pm \XVIdgradRe & \XXVIIIgradRe \pm \XXVIIIdgradRe \\
                       & $\nabla_{\mbox{[Fe/H]}}$ (dex~arcmin$^{-1}$) & \XVIgradamin \pm \XVIdgradamin & \XXVIIIgradamin \pm \XXVIIIdgradamin \\
                       & $\nabla_{\mbox{[Fe/H]}}$ (dex~kpc$^{-1}$) & \XVIgradkpc \pm \XVIdgradkpc & \XXVIIIgradkpc \pm \XXVIIIdgradkpc \\
\hline
Inner Re   & $N$                        & \XVIinnerN                                                       & \XXVIIIinnerN \\
           & $\langle$[Fe/H]$\rangle$ & \XVImeaninnerFeH^{+\XVImeaninnerFeHuerr}_{-\XVImeaninnerFeHlerr} & \XXVIIImeaninnerFeH^{+\XXVIIImeaninnerFeHuerr}_{-\XXVIIImeaninnerFeHlerr} \\ 
           & $\sigma_{\mbox{[Fe/H]}}$   & \XVIdispinnerFeH^{+\XVIdispinnerFeHuerr}_{-\XVIdispinnerFeHlerr} & \XXVIIIdispinnerFeH^{+\XXVIIIdispinnerFeHuerr}_{-\XXVIIIdispinnerFeHlerr} \\ 
           & Skew                       & \XVIskewinnerFeH^{+\XVIskewinnerFeHuerr}_{-\XVIskewinnerFeHlerr} & \XXVIIIskewinnerFeH^{+\XXVIIIskewinnerFeHuerr}_{-\XXVIIIskewinnerFeHlerr} \\ 
           & Kurtosis                   & \XVIkurtinnerFeH^{+\XVIkurtinnerFeHuerr}_{-\XVIkurtinnerFeHlerr} & \XXVIIIkurtinnerFeH^{+\XXVIIIkurtinnerFeHuerr}_{-\XXVIIIkurtinnerFeHlerr} \\ 
\hline
Outer Re   & $N$                        & \XVIouterN                                                       & \XXVIIIouterN \\
           & $\langle$[Fe/H]$\rangle$ & \XVImeanouterFeH^{+\XVImeanouterFeHuerr}_{-\XVImeanouterFeHlerr} & \XXVIIImeanouterFeH^{+\XXVIIImeanouterFeHuerr}_{-\XXVIIImeanouterFeHlerr} \\ 
           & $\sigma_{\mbox{[Fe/H]}}$   & \XVIdispouterFeH^{+\XVIdispouterFeHuerr}_{-\XVIdispouterFeHlerr} & \XXVIIIdispouterFeH^{+\XXVIIIdispouterFeHuerr}_{-\XXVIIIdispouterFeHlerr} \\ 
           & Skew                       & \XVIskewouterFeH^{+\XVIskewouterFeHuerr}_{-\XVIskewouterFeHlerr} & \XXVIIIskewouterFeH^{+\XXVIIIskewouterFeHuerr}_{-\XXVIIIskewouterFeHlerr} \\ 
           & Kurtosis                   & \XVIkurtouterFeH^{+\XVIkurtouterFeHuerr}_{-\XVIkurtouterFeHlerr} & \XXVIIIkurtouterFeH^{+\XXVIIIkurtouterFeHuerr}_{-\XXVIIIkurtouterFeHlerr} \\
\enddata
\tablecomments{Summarizing metallicity properties of And~XVI and And~XXVIII derived from this work, along with the number of stars used to infer each property.} 
\end{deluxetable*}

\section{Discussion} \label{sec:discussion}

\subsection{MDFs: Context and Interpretation}
    
\par Within the M31 satellite system, the most well-sampled stellar spectroscopic measurements have mostly been made in the more massive satellites \citep{vargas2014m31alpha,ho2015,kvasova2024and18}. Due to the difficulty of measuring metallicities for individual stars at the large distance of M31 satellites, most metallicites to-date are measured by stacking low S/N stellar spectra or using estimates from broadband photometry (e.g., \citealt{collins2015m31plane}, \citealt{martin2014m31dwarfspec}).  These studies report mean metallicities from the result of spectral stacking and provide little information on MDFs.

\par Of the few faint M31 dwarfs with individual star spectroscopic metalliciites \citep{kirby2020m31dwarfs}, in terms of stellar mass, And~V ($M_{\star} = 10^{5.8}~M_{\odot}$, 81 stars out to $2.5~R_e$) is closest to And~XXVIII and And~X ($M_{\star} = 10^{5.1}~M_{\odot}$, 21 stars out to $1.5~R_e$) is closest to And~XVI. The mean metallicities of the dwarf galaxies in each respective comparison pair are comparable, and places them within scatter on the universal dwarf galaxy mass-metallicity relation. Like And~XXVIII, And~V also has an MDF with a long MP tail. The MDF of And~X is also symmetric like that of And~XVI, but does not include any stars more metal-rich than $-1.5$. The MDFs of And~V and And~X both have 3 stars below $\mbox{[Fe/H]} < -3.0$. The percentage of EMP stars in And~V and And~XXVIII are similar. And~X has a higher fraction of EMP stars than And~XVI, although the sample size of And~X's MDF is smaller. Owing to their large distances and weak lines, EMP stars in the M31 satellite system have not been extensively identified in detail in the way that they have been in, e.g., MW satellites \citep{starkenburg2013sculptorlowmet}. The EMP candidates we designate in this study will be valuable spectroscopic follow-up targets in the era of ELT spectroscopy.

\subsubsection{And~XXVIII}

\par From the compilation of properties of MW satellites by \citet{kirby2013LZR}, the closest luminosity analogs to And~XXVIII ($L \sim 10^{5.5}~L_{\odot}$) are Ursa Minor ($\langle \mbox{[Fe/H]} \rangle = -2.13$, $\sigma_{\mbox{[Fe/H]}} = 0.43$), Draco ($\langle \mbox{[Fe/H]} \rangle = -1.98$, $\sigma_{\mbox{[Fe/H]}} = 0.42$), and Canes Venatici~I ($\langle \mbox{[Fe/H]} \rangle = -1.91$, $\sigma_{\mbox{[Fe/H]}} = 0.44$). Comparing qualitatively to these MDFs, the MDF of And~XXVIII resembles those of Draco and Canes Venatici~I in that the MDFs of all three galaxies are characterized by long, MP tails. In contrast, the MDF of Ursa Minor is more symmetric, with a slightly longer MR tail. Quantitatively, the mean metallicity of And~XXVIII is comparable to all three galaxies, although its metallicity dispersion is smaller by $\sim0.1$~dex. This smaller dispersion is due to the MP side of the MDF: we assume large uncertainties on our EMP star candidates, which downweighs the most MP stars' contributions to the metallicity dispersion. A re-inference of And~XXVIII's MDF without assuming our adopted systematic uncertainties, which mostly affect the most metal-poor stars, yields $\langle \mbox{[Fe/H]} \rangle = -1.99 \pm 0.05$ and $\sigma_{\mbox{[Fe/H]}} = 0.54 \pm 0.04$.  This suggests that differences in $\sigma$ for the MW satellites and And~XXVIII are not significant. The higher order moments in their MDFs are also comparable. 


\subsubsection{And~XVI}

\par At $L \sim 10^{4.9}~L_{\odot}$, And~XVI is classified as one of the most luminous ultra-faint dwarf galaxies ($L < 10^5~L_{\odot}$; \citealt{simon2019review}) in the LG . Its closest luminosity analog among MW UFDs is Eri~II ($L \sim 10^{4.8}~L_{\odot}$, $\langle \mbox{[Fe/H]} \rangle = -2.63 \pm 0.06$, $\sigma_{\mbox{[Fe/H]}} = 0.26^{+0.08}_{-0.09}$; \citealt{crnojevic2016eriII}, \citealt{li2017eriII}, \citealt{fu2022eriII}). Compared to Eri~II, the MDF of And~XVI is more MR by 0.5~dex. While And~XVI also has an MP tail, it also has a higher fraction of stars more MR than $\mbox{[Fe/H]}=-2.0$, and, subsequently, a higher $\sigma_{\mbox{[Fe/H]}}$. This may be attributable to its extended SFH relative to Eri II, which allows more time for subsequent generations of stars to enrich. The gradual fall-off of And~XVI's MDF suggests that it likely did not experience rapid truncation of SF (e.g., via ram pressure stripping or strong feedback) during the final epochs of its star formation.\footnote{A stark contrast is the MDF of the more massive And~XVIII, which sharply truncates at $\mbox{[Fe/H]}=-1.0$; this cut-off has been attributed to a sudden stop to star formation \citep{kvasova2024and18}.} 

\par The HST CMD-based SFH  of And~XVI shows that it had two major episodes of SFH: one at 13~Gyr ago that formed $\sim50\%$ of its stellar mass before being quenched at a time corresponding to reionization, and then another SF at 8~Gyr that built up the remainder of the galaxy's stellar population (\citealt{weisz2014m31sfhs}, \citealt{monelli2016and16}, \citealt{skillman2017islands}). Given this SFH, a speculative interpretation of its potentially double-peaked MDF is that the MP peak ($\mbox{[Fe/H]}=-2.5$) and MR ($\mbox{[Fe/H]}=-1.7$) peak respectively correspond to its first and second major episodes of star formation. Following this narrative thread, and assuming a present-day $M_{\odot}/L_{\odot}=2$, And~XVI at 13~Gyr ago may have had $M_{\star} \sim 8 \times 10^4~M_{\odot}$, with a mean stellar metallicity of $-2.5$. At this time, its stellar mass and mean metallicity would have been very comparable to the present-day mass of MW UFDs CVn~II, Hya~II, and Eri~II \citep{sacchi2021,fu2023ufds,weisz2023eriII}. 

\par However, the age-metallicity relation (AMR) of And~XVI from the HST CMD modeling measured by \citet{monelli2016and16} is flat and therefore does not suggest metallicity evolution as a function of time. The uncertainties on the metallicities from the AMR encompass the peaks of the MDF, so it's possible that SFH studies may not have the metallicity sensitivity to be able to resolve this enrichment scenario.  Moreover,  \citet{monelli2016and16} only use Solar-scaled stellar models, which may not be accurate for the lower metallicity stars in And~XVI, possibly affecting the metallicity inference.  Obtaining more stellar metallicities through deeper HST CaHK imaging of the current And~XVI field, or HST CaHK and broadband imaging across a different field in the galaxy may be the best way to determine whether the double peaked feature in the MDF is real. 


\subsection{Metallicity Gradients in M31 Dwarf Galaxies} 

\par From \XVINtot~stars spread over 2 $R_e$, we measure a metallicity gradient in And~XVI at 1.5-$\sigma$ significance. The identification of a gradient in And~XVI is interesting, as the extent to which lower-mass galaxies like UFDs can host metallicity gradients, and what the origins of their gradients could be, is still unknown. There are many ongoing efforts to detect stars in the outskirts of MW UFDs, which tend to be even lower in stellar mass than And~XVI (\citealt{waller2023ufdoutskirts}, \citealt{longeard2023hercules}, \citealt{tau2024UFDextended}, \citealt{ou2024hercules}). Among these studies, two have suggested the existence of metallicity gradients in the studied UFDs. In Eri~II, the metallicities of 67 RR Lyrae stars out to $2~R_e$ trace out a gradient of $\sim 0.3~dex~R_e^{-1}$ \citep{martinezvazquez2021}.\footnote{Our own analysis of Eri~II's inner $R_e$ using HST CaHK imaging did not uncover a gradient \citep{fu2022eriII}. The area is small and we did not consider $\alpha$-enhnacements in that analysis, which may contribute to our result.} In the lower mass Tucana~II ($L_{\star} \sim 3 \times 10^{3}~L_{\odot}$), the metallicity difference between the inner regions ($\mbox{[Fe/H]}=-2.6$) and the two MP ($\mbox{[Fe/H]}=-3.0$) stars beyond $2.5~R_e$ forms the basis for the gradient. The differential variations here are comparable to the gradient strength that we measure in And~XVI. 

\par Given the intrinsic difficulty of detecting metallicity gradients in faint galaxies (e.g., few stars available for targeting, coverage and selection effects), we consider the implications of a metallicity gradient in And~XVI. One of its most notable differences from other known UFDs is an extended SFH: MW UFDs universally have old stellar populations of $\sim13~Gyr$ \citep{brown2014sfh,weisz2014sfh,sacchi2021}, as do the majority of M31 UFDs with published SFHs \citep{savino2023}. Simulations have posited that metallicity gradients can form in dwarf galaxies more massive than And~XVI via mechanisms that require an extended SFH: e.g., stellar feedback from young, short-lived stars that preferentially push old, metal-poor stars to a galaxy's outskirts (FIRE simulations; \citealt{elbadry2016gradients, mercado2021gradients}), gas accretion which triggers centrally-concentrated star formation (e.g., \citealt{schroyen2013grads}). The detection of a gradient in And~XVI, which lies at the threshold of the nominal division ($M_{\star} \sim 10^5~M_{\odot}$) between UFDs and classical dwarf galaxies, may suggest that similar physics can be significant in shaping the evolution of UFDs at the mass of And~XVI. As we discuss later in this section, additional theoretical work is necessary to determine the mechanisms for stellar metallicity gradient formation in such low-mass systems.

\par We resolve a strong, highly-significant (4.6-$\sigma$) metallicity gradient in And~XXVIII, where the average metallicity at the center of the galaxy is more MR than stars at $2~R_e$ by 0.9~dex. A more general population gradient in And~XXVIII was also observed by \citet{slater2015spectroscopy} in the different spatial concentrations of its red and blue horizontal branch stars. In addition, stars between 1 and 2 $R_e$ in And~XXVIII have a larger metallicity dispersion than stars in its center, which is also seen in the spatially extended metal-poor population of Ursa Minor \citep{pace2020umi}. Among Local Group dwarf galaxies with population gradients, spatially-compact metal-rich populations and comparatively spatially diffuse metal-poor populations can also have different kinematics (e.g., \citealt{battaglia2006fornax}, \citealt{walker2009massprofiles}, \citealt{kacharov2017}, \citealt{pace2020umi}). The presence or absence of this feature may point to different origins for forming the gradient (e.g., kinematic differences suggestive to dwarf-dwarf mergers; \citealt{taibi2022metgrads}). In this context, And~XXVIII is also an ideal target for spectroscopic studies to obtain kinematic information, enabling a full chemodynamic characterization of the system and its formation history. 

\begin{figure}[h]
    \epsscale{1.2}
    \centering
    \includegraphics[scale=0.43]{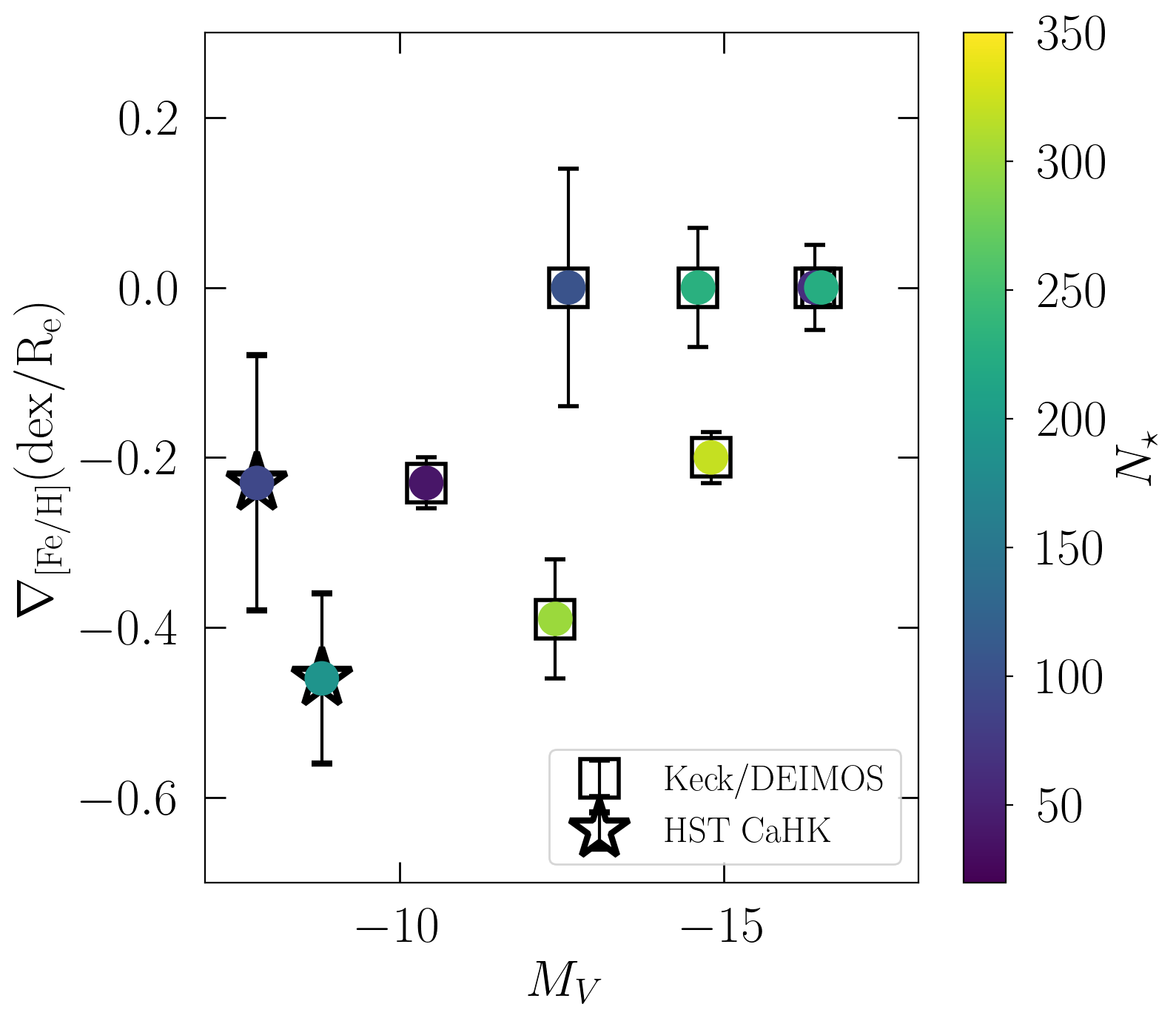}
    \caption{Contextualizing our gradient measurements for And~XVI and And~XXVIII in the current landscape of gradients in M31 dwarfs. Literature gradients for M31 satellites are from the re-analysis of the \citet{ho2015} dataset in \citet[][NGC205, NGC185, NGC 147, And~VII]{taibi2022metgrads}, and \citet[][And~XVIII]{kvasova2024and18}. The galaxies are color-coded by the number of stars used to make the gradients measurement. We have measured gradients in these faint galaxies using a comparable number of stars that Keck/DEIMOS spectroscopy was able to reach in dwarf galaxies several orders of magnitude more luminous.}
    \label{fig:m31gradients}
\end{figure}

\par Metallicity gradients are common among ``classical'' Local Group dwarf galaxies. They have been observed within galaxies spanning a range of kinematic properties, isolated/satellite status, and stellar mass \citep{taibi2022metgrads}. Within the M31 system, however, metallicity gradients have only been measured in three satellites to-date: And~II ($\nabla_{\mbox{[Fe/H]}} = -0.39 \pm 0.07$~dex~$R_e^{-1}$) and NGC~185 ($\nabla_{\mbox{[Fe/H]}} = -0.20 \pm 0.03$~dex~$R_e^{-1}$) from the \citet{vargas2014m31alpha} and \citet{ho2015} dataset that was re-analyzed by \citet{taibi2022metgrads}, and And~XVIII ($\nabla_{\mbox{[Fe/H]}} = -0.23 \pm 0.03$~dex~$R_e$; \citealt{kvasova2024and18}).\footnote{Despite measuring the gradient from only 38 stars, the reported uncertainties in And~XVIII are smaller than from gradient measurements made with hundreds of stars as in the case of the previous work. \citet{kvasova2024and18}, as well as \citet{vargas2014m31alpha} and \citet{ho2015}, used a least-squares method to infer their gradients, which results in smaller gradient uncertainties due to assuming known Gaussian variance on individual stellar metallicities \citep{hogg2010datarecipes}. For a more direct comparison with our inference approach, we instead adopt the \citet{taibi2022metgrads} re-analysis where available, as we find from \citet{fu2024tucana} that their Gaussian process regression method produces comparable gradient measurements to ours.} 

\par In Figure \ref{fig:m31gradients}, we emphasize the novelty of our measurements from this paper within the landscape of gradient measurements in M31 dwarfs. In more luminous M31 dwarf galaxies, Keck/DEIMOS spectroscopy has been used to constrain metallicity gradients from measurements of hundreds of stars. In fainter M31 satellites, there simply aren't enough bright stars to construct comparably well-populated MDFs using spectra. The CaHK-based MDFs in this paper, however, can provide large samples of metallicities, enabling the construction of MDFs that are as robust as the more luminous M31 satellites.

\par Metallicity gradients are one of the few observationally accessible signatures of baryonic processes that shape both luminous and dark matter. For example, stellar feedback has long been invoked as a mechanism for transforming initially cuspy dark matter haloes into cores (\citealt{navarro1996outflowsdm}, \citealt{governato2010outflows}, \citealt{garrisonkimmel2013feedbacktbtf}, \citealt{madau2014coreheating}). Through this mechanism, cored DM profiles would therefore be consistent with a $\Lambda$CDM universe, providing resolution to the long-standing ``core-cusp'' problem. As one example in the contemporary literature, the FIRE-2 simulations of isolated dwarf galaxies predict a relationship between gradient strength and mean stellar age in dwarf galaxies \citep{mercado2021gradients} that would arise as a result of these feedback processes that also produce dark matter cores.

\begin{figure*}
    \epsscale{1.2}
    \centering
    \includegraphics[scale=0.45]{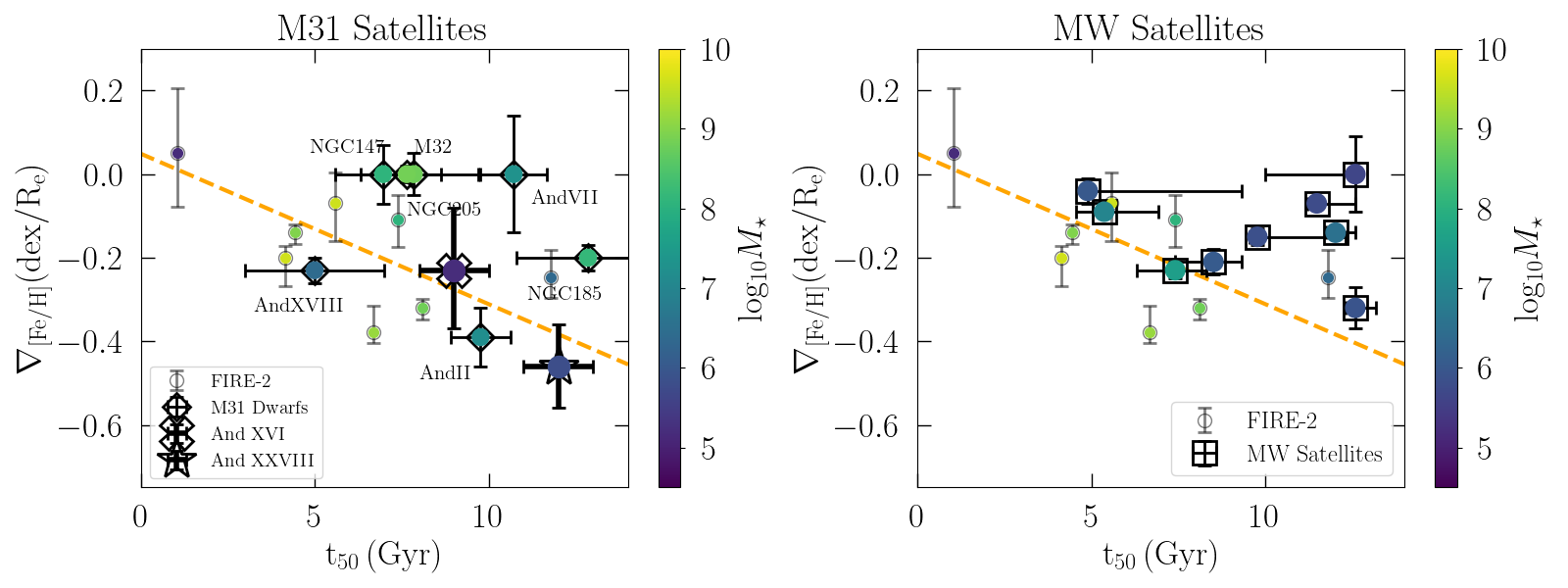}
    \caption{Comparing M31 and MW satellite metallicity gradients to predictions from the FIRE-2 simulations. \textbf{(Left)} Presenting gradients of M31 satellites relative to the age-metallicity gradient strength relation from FIRE-2 simulations \citep{mercado2021gradients}. All galaxies, simulated and observed, are color-coded by their stellar masses. The linear form of the relation is orange, and light circles denote the publicly-available FIRE-2 simulated galaxies. Diamonds are M31 satellites from the literature, and the 'X' and star points are respectively And~XVI and And~XXVIII. Our dwarf galaxy gradient measurements trace out the age-gradient relation. \textbf{(Right)} Presenting gradients of MW satellites as compared to the FIRE-2 simulations. The simulated relations are presented in the same fashion as the previous panel. Gradient measurements are sourced from \citet{taibi2022metgrads}. The MW satellites show large scatter about the age-gradient relation.}
    \label{fig:mwm31gradientssims}
\end{figure*}

\par We show this relation in Figure \ref{fig:mwm31gradientssims} and compare it to data. In the left panel, we plot the linear form of the relation fit to \citet{mercado2021gradients} simulations, properties of the publicly-available simulated FIRE dwarfs, and the gradient measurements of M31 dwarfs, including those derived in this work. With the exception of And~XVIII \citep{kacharov2017}, the mean galaxy ages ($t_{50}$) are taken from Savino et al. (in prep) which use deep CMDs from the HST survey of M31 satellites to measure SFHs, and includes re-analyses of CMDs from the literature (\citealt{geha2015n147185}, \citealt{skillman2017islands}). 

In the right panel, we compare the same relation against gradient measurements of MW satellites from the \citet{taibi2022metgrads} analysis, with SFHs from \citet{weisz2014sfh}. All dwarfs, simulated and observed, are color-coded by their stellar mass. One notable caveat to data interpretation here is that the area coverage of the SFH measurements and the spectroscopic gradient measurements are often different, with the gradient measurements spanning a larger area over the dwarf due to the nature of spectroscopic targeting. 

\par Our measurements of And~XVI and And~XXVIII appear to affirm the theoretical relation quite well. This is additionally noteworthy given that the relation was derived from simulations of isolated dwarfs, not satellites. This could be because the bulk of the metallicity information was in place at early times before many of these galaxies were bound satellites. The rest of the M31 satellites, as well as the MW satellites, appear to have large scatter about this relation. In particular, there are no galaxies with intermediate $t_{50}$ that have strong gradients. The scatter in gradient strength is largest among galaxies with older stellar populations, ranging from non-detections to gradients as strong as $-0.4$~dex~$R_e^{-1}$. One simple, qualitative interpretation of this picture is that old galaxies, with metallicity gradients in place in accordance to the age-gradient relation, and had their gradients flattened through close passage(s) with a more massive host that preferentially removes their MP outskirts.\footnote{Simulations of tidal stripping in dwarf galaxies show that the outskirts of the galaxy are removed first \citep[e.g., ][]{penarrubia2008tidalevLG,errani2022dwarftideslcdm}.} The extent to which And~XVI and And~XXVIII can be described by this scenario is unclear: both currently lie beyond the virial radius of M31, and their orbital histories about M31 are not constrained due to absence of proper motions. Detailed orbital histories of LG dwarf galaxies \citep{patel2020,pace2022orbits, battaglia2022orbitalprops,bennet2023PMsorbits} will be key to interpreting the various astrophysical origins of these observed gradients.

\par The theoretical age-gradient relation is compelling for its potential as an observational diagnostic of dark matter core formation in the dwarf galaxy population. But, its explanatory power needs refinement, given that environmental processes can also set metallicity gradients (\citealt{marcolini2008}, \citealt{sales2010tidal}, \citealt{benitezllambay2016}, \citealt{hausammann2019}). Comparison between observation and theory also have additional caveats, in part due to ongoing challenges in simulating low-mass galaxies to high fidelity. The faintest publicly-available FIRE-2 dwarf galaxy used to infer the age-gradient relation has $M_{\star} \sim 10^5~M_{\odot}$ and $t_{50}$ of $1$~Gyr; these properties are not well-matched to any LG dwarf galaxy regardless of environment, as galaxies in this mass range are dominated largely by old ($<10$~Gyr), or in some cases intermediate ($\sim6$~Gyr) age stellar populations \citep[e.g.,][]{weisz2014sfh, gallart2015, skillman2017islands}. 

\par Notably, theoretical work, including the same simulations positing the age-gradient relation, have not identified galaxy stellar mass as a strong driving factor in gradient formation \citep{schroyen2013grads,revaz2018,mercado2021gradients}. This runs counter-intuitive to observations, which show that more massive dwarf galaxies tend to have extended SFHs, younger stars, and lower $t_{50}$ values \citep{weisz2014sfh}. If stellar age indeed traces gradient strength, then this relation should also have an underlying stellar mass dependence. Given the near-uniformly old stellar ages of faint dwarf galaxies, particularly UFDs, \citep{brown2014sfh,weisz2014sfh,gallart2021eriii,sacchi2021,simon2023retii}, taking the age-gradient relation at face value would imply that nearly all galaxies with $M_{\star}<10^5~M_{\odot}$ should have strong metallicity gradients. The current empirical landscape of faint dwarf galaxy studies does not support this prediction, although obtaining high-fidelity gradient measurements for a large sample of UFDs has also been challenging. 

\par Still, in the regime of classical dwarf galaxies, it is notable that the population-level behavior of metallicity gradients can be interpreted as the consequence of stellar feedback-driven breathing modes that form dark matter cores (\citealt{stinson2007breathing}, \citealt{elbadry2016gradients}, \citealt{mercado2021gradients}). A natural question is how low in galaxy stellar or halo mass this framework can still apply, and relatedly, whether metallicity gradients in UFDs imply core formation at all. Indeed, whether UFDs host cores is still an open question: while simulations have suggested that stellar feedback is insufficient to sculpt cores in UFDs (e.g., \citealt{munshi2021}, \citealt{azartashnamin2024feedback}), observations of the Eri~II UFD suggest strong feedback during star formation \citep{gallart2021eriii,sandford2022eriII} and a marginally off-center star cluster that is expected to survive to $z=0$ only if the galaxy has a cored DM profile \citep{amorisco2017clusterdmcusps,simon2021EriII,weisz2023eriII}. 

\par The central mass profile of UFDs, and their reconciliability with small-scale predictions under CDM, remains an outstanding question \citep{bullock2017dmreview}. On the observational side, gradient measurements in UFDs across a range of environments will be important for disentangling secular and external impacts on gradient formation. In particular, isolated low-mass galaxies ($M_{\star} < 10^6~M_{\odot}$) with a range of SFHs such as Leo P, Tucana B, and Pavo -- that are not expected to have interacted with a massive host throughout their lifetime -- are important targets as a comparison foil for studies of MW and M31 satellites \citep{mcquinn2015leop,sand2022tucanab,jones2023pavo}. By nature, these high-priority targets also tend to lie at the farthest distances, making spectroscopic studies to obtain adequately sampled stellar metallicity measurements, or stellar velocities for dynamical mass profile studies, especially challenging. As our team's studies thus far have demonstrated, the stellar metallicity gradient measurements enabled by HST CaHK imaging can provide additional constraining power for various gradient formation mechanisms. On the theoretical side, detailed studies of metallicity gradients in high-fidelity simulations of dwarf galaxies, particularly in the UFD class, will be necessary for refining the interpretive lenses to bring to these observations.

\section{Summary} 
\label{sec:conclusion}

\par In this paper, we present the first well-sampled MDFs for And~XVI ($L = 10^{4.9}~L_{\odot}$) and And~XXVIII ($L = 10^{5.5}~L_{\odot}$) based on individual stellar metallicities measured using HST Ca H\&K narrow-band imaging. With a single HST pointing in each galaxy, we measure stellar metallicities with sample sizes and to precisions that ground-based Keck/DEIMOS spectroscopy has been able to attain only in M31 satellite galaxies that are hundreds or thousands of times brighter. We summarize the key results enabled by this novel dataset: 

\begin{itemize}
    \item From \XVINtot~stars in And~XVI, we measure $\langle \mbox{[Fe/H]} \rangle = \XVImeanFeH^{+\XVImeanFeHuerr}_{-\XVImeanFeHlerr}$, in good agreement with previous measurements from stacked stellar spectra. We also resolve $\sigma_{\mbox{[Fe/H]}}=\XVIdispFeH^{+\XVIdispFeHuerr}_{-\XVIdispFeHlerr}$. Though its MDF is well-described by a Gaussian statistically, visually it is also doubly-peaked. We also measure a metallicity gradient ($\nabla_{\mbox{[Fe/H]}}= \XVIgradRe \pm \XVIdgradRe$~dex~$R_e^{-1}$) at a significance of 1.5~$\sigma$. 
    \item Compared to MW UFDs of similar luminosity, And~XVI is more metal-rich by $\sim0.5$~dex. This property, alongside the putative double-peaked structure and the metallicity gradient, may result from its extended SFH, which is unusual among the currently-known UFD population.
    \item From \XXVIIINtot~stars in And~XXVIII, we measure $\langle \mbox{[Fe/H]} \rangle = \XXVIIImeanFeH^{+\XXVIIImeanFeHuerr}_{-\XXVIIImeanFeHlerr}$, $\sigma_{\mbox{[Fe/H]}}=\XXVIIIdispFeH^{+\XXVIIIdispFeHuerr}_{-\XXVIIIdispFeHlerr}$, and a strong metallicity gradient within $2~R_e$: $\nabla_{\mbox{[Fe/H]}}= \XXVIIIgradRe \pm \XXVIIIdgradRe$~dex~$R_e^{-1}$.
    \item And~XXVIII is more metal-rich than And~XVI despite its shorter star formation history; these results affirm that stellar mass is a fundamental property setting the overall enrichment level of a galaxy \citep[e.g., ][]{lee2006MZR,calura2009mzr,mannucci2010masssfrmets,kirby2013LZR}. With forthcoming chemical evolution studies, we will present a detailed characterization of the baryon cycle in galaxies of different masses with different SFHs. 
    \item The metallicity gradient measurements for these galaxies follow the age-gradient relation predicted in the FIRE-2 simulations \citep{mercado2021gradients}, as the observational consequence of gradients formed from the same stellar feedback hypothesized to form dark matter cores at the center of dwarf galaxy halos \citep{stinson2007breathing,elbadry2016gradients}. 
\end{itemize}

\par As the only other current galactic ecosystem for which we can conduct resolved stellar population studies, M31 and its satellite system offer a necessary comparison point to the MW and its satellites. In particular, their study is critical to ongoing efforts to understand the impact of environment on dwarf galaxy evolution, the extent to which reionization has a universal impact on low-mass halos, and imprints of star formation baryonic feedback processes that can reconfigure luminous and dark matter within a galaxy. As our team has demonstrated in this paper and elsewhere, photometric metallicity techniques are an efficient way to obtain detailed pictures of stellar metallicities in distant galaxies, and will be highly complementary with future resolved stellar spectroscopy campaigns with JWST and ELT in the coming years. 

\begin{acknowledgements}

\par SWF acknowledges support from a Paul \& Daisy Soros Fellowship and from the NSFGRFP under grants DGE 1752814 and DGE 2146752. SWF, DRW, and AS acknowledge support from HST-GO-15901, HST-GO-15902, HST-AR-16159, HST-GO-16226, and HST-GO-16686 from the Space Telescope Science Institute, which is operated by AURA, Inc., under NASA contract NAS5-26555. ES acknowledges funding through VIDI grant "Pushing Galactic Archaeology to its limits" (with project number VI.Vidi.193.093) which is funded by the Dutch Research Council (NWO). This research has been partially funded from a Spinoza award by NWO (SPI 78-411). MLMC acknowledges support from STFC grants ST/Y002857/1 and ST/Y002865/1. MBK acknowledges support from NSF CAREER award AST-1752913, NSF grants AST-1910346 and AST-2108962, NASA grant 80NSSC22K0827, and HST-AR-15809, HST-GO-15658, HST-GO-15902, HST-AR-16159, HST-AR-17028, and HST-AR-17043 from the Space Telescope Science Institute, which is operated by AURA, Inc., under NASA contract NAS5-26555. 

\par This research used the Savio computational cluster resource provided by the Berkeley Research Computing program at the University of California, Berkeley (supported by the UC Berkeley Chancellor, Vice Chancellor for Research, and Chief Information Officer). Some/all of the data presented in this paper were obtained from the Mikulski Archive for Space Telescopes (MAST) at the Space Telescope Science Institute. The specific observations analyzed can be accessed via the following DOI link: \href{http://dx.doi.org/10.17909/bzcz-r494}{10.17909/bzcz-r494}. STScI is operated by the Association of Universities for Research in Astronomy, Inc., under NASA contract NAS5–26555. Support to MAST for these data is provided by the NASA Office of Space Science via grant NAG5–7584 and by other grants and contracts.

\end{acknowledgements}

\appendix

\section{Table of Metallicity Measurements}
\label{sec:metappendix}

\par In this section, we present Table \ref{tab:indivmeasurements}, which contains all of the individual stellar metallicity measurements for And~XVI and And~XXVIII. In Table \ref{tab:EMPstars}, we present all of the EMP star candidates in the two galaxies.

\begin{deluxetable*}{ccCCCCCCCCc}
\tabletypesize{\scriptsize}
\tablecaption{Individual Stellar Metallicities}
\label{tab:indivmeasurements}
\tablehead{ \colhead{Galaxy} & \colhead{Star} & \colhead{R.A.} & \colhead{Decl.} & \colhead{F814W} & \colhead{F475W} & \colhead{F395N} & \colhead{VI} & \colhead{CaHK} & \colhead{[Fe/H]} & \colhead{Notes} }
\startdata
 AndXVI &  0 & 14.883476 & 32.381190 & 20.064 \pm 0.001 & 22.175 \pm 0.002 & 24.056 \pm 0.037 & 2.111 \pm 0.002 & -1.286 \pm 0.037 &      -1.77^{+0.10}_{-0.10} \pm 0.2 \mbox{(syst.)}  &    Constrained \\
 AndXVI &  1 & 14.880947 & 32.381578 & 20.291 \pm 0.001 & 22.333 \pm 0.002 & 23.929 \pm 0.034 & 2.042 \pm 0.002 & -1.467 \pm 0.034 &      -2.19^{+0.08}_{-0.07} \pm 0.2 \mbox{(syst.)}  &    Constrained \\
 AndXVI &  2 & 14.877938 & 32.376751 & 20.332 \pm 0.001 & 22.289 \pm 0.002 & 23.719 \pm 0.034 & 1.957 \pm 0.002 & -1.506 \pm 0.034 &      -2.41^{+0.09}_{-0.08} \pm 0.2 \mbox{(syst.)}  &    Constrained \\
 AndXVI &  3 & 14.858829 & 32.384565 & 20.414 \pm 0.001 & 22.381 \pm 0.002 & 23.713 \pm 0.024 & 1.967 \pm 0.002 & -1.618 \pm 0.024 &     < -2.43 &           Ulim \\
 AndXVI &  4 & 14.894606 & 32.384091 & 20.487 \pm 0.001 & 22.420 \pm 0.002 & 23.788 \pm 0.032 & 1.933 \pm 0.002 & -1.532 \pm 0.032 &      -2.46^{+0.10}_{-0.05} \pm 0.2 \mbox{(syst.)}  &    Constrained \\
 AndXVI &  5 & 14.898027 & 32.374751 & 20.652 \pm 0.001 & 22.532 \pm 0.002 & 23.963 \pm 0.034 & 1.880 \pm 0.002 & -1.389 \pm 0.034 &      -2.13^{+0.14}_{-0.16} \pm 0.2 \mbox{(syst.)}  &    Constrained \\
 AndXVI &  6 & 14.872155 & 32.388600 & 20.995 \pm 0.001 & 22.788 \pm 0.003 & 24.059 \pm 0.037 & 1.793 \pm 0.003 & -1.418 \pm 0.037 &      -2.33^{+0.17}_{-0.25} \pm 0.2 \mbox{(syst.)}  &    Constrained \\
 AndXVI &  7 & 14.878955 & 32.384221 & 21.030 \pm 0.002 & 22.831 \pm 0.003 & 24.135 \pm 0.035 & 1.801 \pm 0.004 & -1.397 \pm 0.035 &      -2.21^{+0.13}_{-0.26} \pm 0.2 \mbox{(syst.)}  &    Constrained \\
 AndXVI &  9 & 14.906816 & 32.380212 & 21.117 \pm 0.002 & 22.860 \pm 0.003 & 24.134 \pm 0.031 & 1.743 \pm 0.004 & -1.340 \pm 0.031 &      -2.07^{+0.14}_{-0.13} \pm 0.2 \mbox{(syst.)}  &    Constrained \\
 AndXVI & 10 & 14.864351 & 32.377186 & 21.269 \pm 0.002 & 23.013 \pm 0.003 & 24.206 \pm 0.036 & 1.744 \pm 0.004 & -1.423 \pm 0.036 &      -2.37^{+0.19}_{-0.25} \pm 0.2 \mbox{(syst.)}  &    Constrained \\
\enddata
\tablecomments{Measurements for all of the stars analyzed in this work. We present a portion of the table here for form and content, and will make the full version electronically available.}
\end{deluxetable*}

\begin{deluxetable*}{ccCCCCCCCCc}
\tabletypesize{\scriptsize}
\tablecaption{Candidate EMP Stars}
\label{tab:EMPstars}
\tablehead{ \colhead{Galaxy} & \colhead{Star} & \colhead{R.A.} & \colhead{Decl.} & \colhead{F814W} & \colhead{F475W} & \colhead{F395N} & \colhead{VI} & \colhead{CaHK} & \colhead{[Fe/H]} }
\startdata
    AndXVI &  26 &  14.867180 & 32.379153 & 21.573 \pm 0.002 & 23.239 \pm 0.004 & 24.155 \pm 0.034 &  1.666 \pm 0.004 & -1.583 \pm 0.034 &  -3.41^{+0.33}_{-0.39} \pm 0.5 \mbox{(syst.)} \\
    AndXVI &  30 &  14.879237 & 32.388930 & 21.763 \pm 0.002 & 23.419 \pm 0.003 & 24.400 \pm 0.040 &  1.656 \pm 0.004 & -1.503 \pm 0.040 &  -3.05^{+0.41}_{-0.32} \pm 0.5 \mbox{(syst.)} \\
    AndXVI &  37 &  14.864885 & 32.388081 & 22.062 \pm 0.003 & 23.684 \pm 0.004 & 24.633 \pm 0.044 &  1.622 \pm 0.005 & -1.484 \pm 0.044 &  -3.09^{+0.33}_{-0.41} \pm 0.5 \mbox{(syst.)} \\
    AndXVI &  50 &  14.864719 & 32.378416 & 22.399 \pm 0.003 & 23.935 \pm 0.005 & 24.800 \pm 0.051 &  1.536 \pm 0.006 & -1.439 \pm 0.051 &  -3.05^{+0.51}_{-0.51} \pm 0.5 \mbox{(syst.)} \\
    AndXVI &  80 &  14.863572 & 32.377904 & 22.892 \pm 0.004 & 24.352 \pm 0.006 & 25.143 \pm 0.067 &  1.460 \pm 0.007 & -1.399 \pm 0.067 &  -3.08^{+0.58}_{-0.58} \pm 0.5 \mbox{(syst.)} \\
    AndXVI &  89 &  14.909581 & 32.370058 & 23.232 \pm 0.005 & 24.659 \pm 0.006 & 25.435 \pm 0.067 &  1.427 \pm 0.008 & -1.364 \pm 0.067 &  -3.10^{+0.59}_{-0.59} \pm 0.5 \mbox{(syst.)} \\
    AndXVI &  90 &  14.864426 & 32.384401 & 23.271 \pm 0.006 & 24.698 \pm 0.010 & 25.470 \pm 0.076 &  1.427 \pm 0.012 & -1.369 \pm 0.077 &  -3.16^{+0.62}_{-0.54} \pm 0.5 \mbox{(syst.)} \\
 AndXXVIII &   4 & 338.177742 & 31.228019 & 20.673 \pm 0.001 & 22.960 \pm 0.003 & 24.602 \pm 0.043 &  2.287 \pm 0.003 & -1.789 \pm 0.043 & <-4.00 \\
 AndXXVIII &  21 & 338.151042 & 31.219175 & 21.171 \pm 0.001 & 23.229 \pm 0.002 & 24.681 \pm 0.047 &  2.058 \pm 0.002 & -1.635 \pm 0.047 & <-4.00 \\
 AndXXVIII &  50 & 338.186547 & 31.203947 & 21.456 \pm 0.002 & 23.386 \pm 0.003 & 24.586 \pm 0.044 &  1.930 \pm 0.004 & -1.695 \pm 0.044 & <-4.00 \\
 AndXXVIII &  51 & 338.159931 & 31.215765 & 21.477 \pm 0.001 & 23.416 \pm 0.003 & 24.721 \pm 0.056 &  1.939 \pm 0.003 & -1.604 \pm 0.056 & <-4.00 \\
 AndXXVIII & 119 & 338.153888 & 31.224333 & 22.311 \pm 0.002 & 24.045 \pm 0.004 & 24.995 \pm 0.048 &  1.734 \pm 0.004 & -1.651 \pm 0.048 &  -3.50^{+0.38}_{-0.23} \pm 0.5 \mbox{(syst.)} \\
 AndXXVIII & 129 & 338.172690 & 31.217700 & 22.487 \pm 0.003 & 24.144 \pm 0.005 & 25.132 \pm 0.088 &  1.657 \pm 0.006 & -1.497 \pm 0.088 &  -3.17^{+0.48}_{-0.48} \pm 0.5 \mbox{(syst.)} \\
 AndXXVIII & 134 & 338.155764 & 31.225169 & 22.452 \pm 0.002 & 24.150 \pm 0.005 & 25.218 \pm 0.056 &  1.698 \pm 0.005 & -1.479 \pm 0.056 &  -3.18^{+0.48}_{-0.47} \pm 0.5 \mbox{(syst.)} \\
 AndXXVIII & 145 & 338.185173 & 31.188016 & 22.636 \pm 0.003 & 24.322 \pm 0.005 & 25.314 \pm 0.074 &  1.686 \pm 0.006 & -1.537 \pm 0.074 &  -3.47^{+0.52}_{-0.36} \pm 0.5 \mbox{(syst.)} \\
 AndXXVIII & 157 & 338.189276 & 31.199485 & 22.746 \pm 0.003 & 24.316 \pm 0.005 & 25.221 \pm 0.066 &  1.570 \pm 0.006 & -1.450 \pm 0.066 &  -3.11^{+0.48}_{-0.47} \pm 0.5 \mbox{(syst.)} \\
 AndXXVIII & 174 & 338.142529 & 31.212665 & 23.016 \pm 0.004 & 24.598 \pm 0.006 & 25.506 \pm 0.073 &  1.582 \pm 0.007 & -1.465 \pm 0.073 &  -3.32^{+0.56}_{-0.46} \pm 0.5 \mbox{(syst.)} \\
 AndXXVIII & 192 & 338.151661 & 31.218511 & 23.380 \pm 0.004 & 24.867 \pm 0.006 & 25.770 \pm 0.098 &  1.487 \pm 0.007 & -1.328 \pm 0.098 &  -3.05^{+0.62}_{-0.61} \pm 0.5 \mbox{(syst.)} \\
 AndXXVIII & 196 & 338.148343 & 31.212334 & 23.508 \pm 0.005 & 25.083 \pm 0.008 & 26.033 \pm 0.101 &  1.575 \pm 0.009 & -1.412 \pm 0.101 &  -3.26^{+0.47}_{-0.49} \pm 0.5 \mbox{(syst.)} \\
\enddata
\tablecomments{All of our identified EMP candidates in And~XVI and And~XXVIII.}
\end{deluxetable*}

\bibliography{bibliography}{}
\bibliographystyle{aasjournal}



\end{document}